\documentclass[aps,prbbib,floatfix,floats,twocolumn,preprintnumbers]{revtex4}
\usepackage{graphicx,latexsym}
\usepackage{epsfig}
\usepackage{amsmath}
\usepackage{color}

\newcommand{\be}{\begin{equation}}
\newcommand{\ee}{\end{equation}}
\newcommand{\bea}{\begin{eqnarray}}
\newcommand{\eea}{\end{eqnarray}}

\begin{document}

\title{Hund and anti-Hund rules in circular molecules}

\author{M. Ni\c t\u a}
\address{ National Institute of Materials Physics,POB MG-7, 77125, Bucharest-Magurele, Romania}
\author{M. \c Tolea}
\address{ National Institute of Materials Physics,POB MG-7, 77125, Bucharest-Magurele, Romania}
\author{D. C. Marinescu}
\address{ Department of Physics and Astronomy, Clemson University, Clemson, South Carolina 29634, USA}
\author{A. Manolescu}
\address{ School of Science and Engineering, Reykjavik University, Menntavegur 1, IS-101 Reykjavik, Iceland}

\begin{abstract}
We study the validity of Hund's first rule for the spin multiplicity in circular molecules
 - made of real or artificial atoms such as quantum dots - by considering a perturbative approach in the Coulomb interaction in the extended Hubbard model with both on-site and long-range interactions. In this approximation, we show that an anti-Hund rule {\it always} defines the ground state in a molecule with $4N$ atoms at half-filling.
In all other cases (i.e. number of atoms {\it not} multiple of four, or a $4N$ molecule away from half-filling) both the singlet and the triplet outcomes are possible, as determined {primarily} by the total number of electrons in the system. In some instances, the Hund rule is always obeyed and the triplet ground state is realized {\it mathematically} for any values of the on-site and long range interactions, while for other filling situations the singlet is also possible but only if the long-range interactions exceed a certain threshold, relatively to the on-site interaction.
\end{abstract}

\maketitle

\section{Introduction}\label{introducere}

The study of experimental atomic spectra in connection with the theoretical behaviour of quantum particles
led Hund to formulate
the rules that express the spin and angular momenta of many electron atoms.
Hund's first rule (HFR) states that for a given electronic configuration, corresponding to incompletely occupied outer orbitals, the state with the maximum spin (i.e. multiplicity) is the ground state \cite{Hund1925, Kutzelnigg1996}.

A textbook example of HFR confirmation is the
electron filling of the three degenerate $p$ orbitals in an atomic sub-shell
($p_x$, $p_y$, $p_z$),
when the triplet, rather than the singlet, configuration is realized, as it happens in the C atom. This was explained first by Slater \cite{Slater1929} who considered the antisymmetric nature of the electronic wave function that generates a higher value for the Coulomb repulsion in the quantum state with the lower spin value. More recent
theoretical analyzes were developed in Refs.~\cite{Kutzelnigg1996, Grynova2015}.

Hund rule's relevance has long exceeded the boundaries of atomic physics where it was first formulated,
over the years being investigated in many other systems such as quantum dots
\cite{Tarucha1996, Koskinen1997, Steffens1998, Partoens2000, Matagne2002, He2005,
Sajeev2008, Sako2010, Sako2012, Katriel2012, Schroter2013, Hofmann2016},
artificial molecules created by quantum dots
\cite{Reimann2002, Ota2005, Korkusinski2007, Florens2011},
metal clusters \cite{Kumar2001, Yoshida2015, Sen2016, Reber2017},
bipartite lattices \cite{Sheng2013, Tolea2016}, ultrathin films \cite{Belabbes2016}
new carbon systems \cite{Zhang2012, Jiang2016}
or even in optical lattices \cite{Ho2006, Karkkainen2007}.

A further understanding of the physical mechanism behind HFR is also offered by the study of
 physical systems where the rule is reversed, i. e. the two highest energy
electrons form a singlet rather than a triplet. Such a situation is known to exist, for example, in quantum dots, where the zero spin ground state is associated with
 a spin density wave \cite{Koskinen1997}, or in
 artificial molecules, when the increase of level splitting  overcame
the exchange energy gain by parallel spin alignment \cite{Partoens2000}.
In semiconductor artificial atoms under magnetic field, the Hund rule violation is noticed in connection
with changes in the ground state symmetry \cite{Destefani2004}, while
in quadratically confining quantum dots is related to the modification of the localization properties
of some singlet states \cite{Schroter2013}.

Other exceptions to Hund's rule, in close relation with the phenomenology studied in this paper,
are known to exist in physical systems that exhibit, as a common feature,
degenerate, non-overlapping single-particle states in the mid-spectrum of the electronic
Hamiltonian. Such states, that do not have any common sites around the ring, are called disjoint orbitals and have been identified in
ring-like molecules  \cite{Borden1975, Borden1994, Kollmar1978, Zilberg1998},
graphene nanoflakes \cite{Sheng2013} and small Lieb lattices \cite{Tolea2016}.

The classic example concerning HFR validity in molecules is given by a four-atom molecule such as square cyclobutadiene, for which
the H\" uckel model gives an energy spectrum with four states, two of them being the disjoint orbitals in the middle \cite{Borden1994}.
The model is equivalent with a quadruple quantum dot molecules as described in \cite{Thalineau2012, Ozfidan2013}.
Using a four-electron wave function it was shown that
the singlet state has a lower energy on account of Coulomb correlations associated with single particle excitations
which are absent in the triplet state \cite{Borden1975}.
The quantitative calculation of this result was developed on the base of
the spin polarization phenomena, where within a self-consistent field approximation, the Brillouin theorem
specifies which of the transitions between the many particle spin state are canceled out.
When performed for $C_4H_4$ and $H_4$ molecules in the second order of interaction
this algorithm yields negative singlet-triplet energy gap in a violation of the HFR \cite{Kollmar1978}.

In this paper we discuss the spin properties of circular molecules with an arbitrary number of atoms whose one-particle spectrum, in general,
is composed from a ladder of degenerate electronic states \cite{Salem1966, Jones1979, Siga2005, Mercero2015}. This property recommends them as adequate physical systems
to investigate the  HFR applicability, motivating the significant number of previous studies, as briefly described above.

If eigenvectors of the noninteracting Hamiltonian are used to construct the many-particle states of the interacting system,
the Brillouin theorem { as in Ref.\,\cite{Kollmar1978}} is not applicable anymore,
but the transition probability can be analytically investigated
as done for the Hubbard model in Ref.\,\cite{Tolea2016} where the negative singlet-triplet gap for an octagon molecule was
shown to result from the two mid-spectrum disjoint orbitals and from the electron-hole symmetry of the spectrum.

%

{As a technical detail,}  we consider the extended Hubbard model for the general case of a circular molecule
by including {also} a long-range interaction potential, as described in Section\,II.
The single-particle spectrum which consists of a ladder of double degenerate states is discussed in Section\,III.
The particular case of the molecules with $4N$ atoms which present a pair of degenerate non-overlapping levels at mid-spectrum is emphasized.
Our main formal results are given in Section\,IV, and then applied to some particular cases of interest in Section\,V. Section\,VI concludes the paper.

\section{Circular molecule and the interacting potential}

In a tight-binding approximation, we describe a circular molecule composed of $N_s$ sites (either real atoms or artificial ones, such as quantum dots) occupied by $N_e$ electrons that interact through a long range Coulomb interaction by the Hamiltonian \cite{Mahan},
\begin{eqnarray}\label{hamiltonian}
\hat H & = & -t\sum_{n} ( c_{n+1}^\dagger c_{n} + c_{n}^\dagger c_{n+1} ) \nonumber\\
& + &
\frac{1}{2}\sum_{n,m,\sigma,\sigma'} V_{nm}  c_{n\sigma}^\dagger  c_{m\sigma'}^\dagger c_{m,\sigma'} c_{n\sigma},
\end{eqnarray}
where $c_{n\sigma}^\dagger$ and $c_{n\sigma}$ are the creation and annihilation operators for an electron state of spin
 $\sigma = \pm 1/2$ at location $n= {1,\cdots,N_s}$. Every site $n$ can host a maximum of two
electrons, of opposite spins.

\begin{figure}
\centering
\includegraphics[scale=0.7]{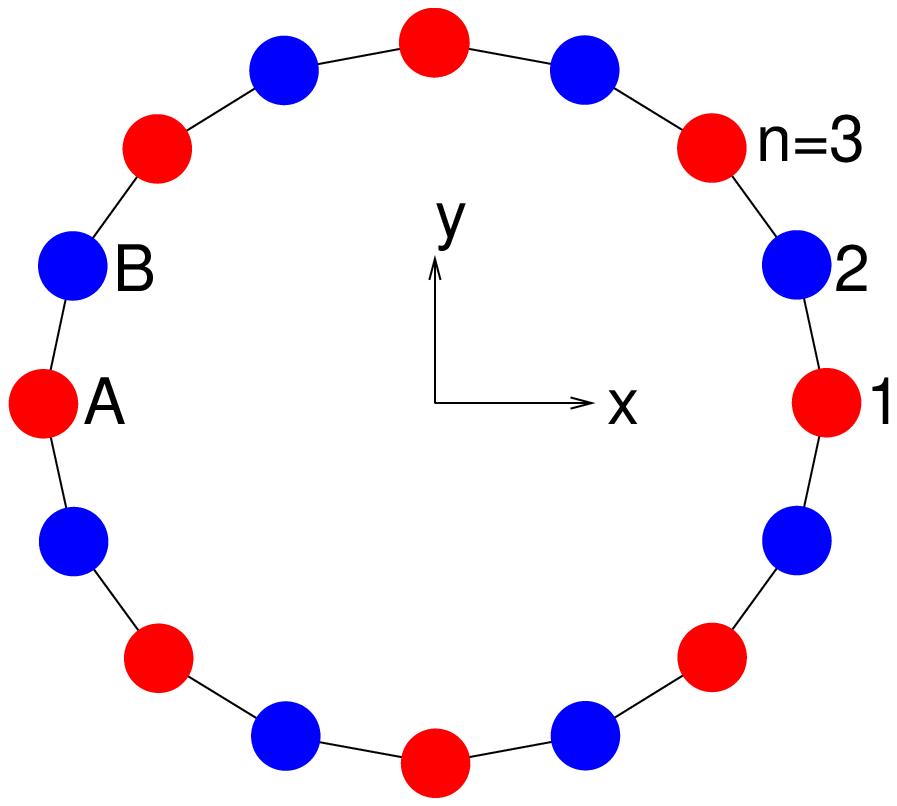}
\includegraphics[scale=0.7]{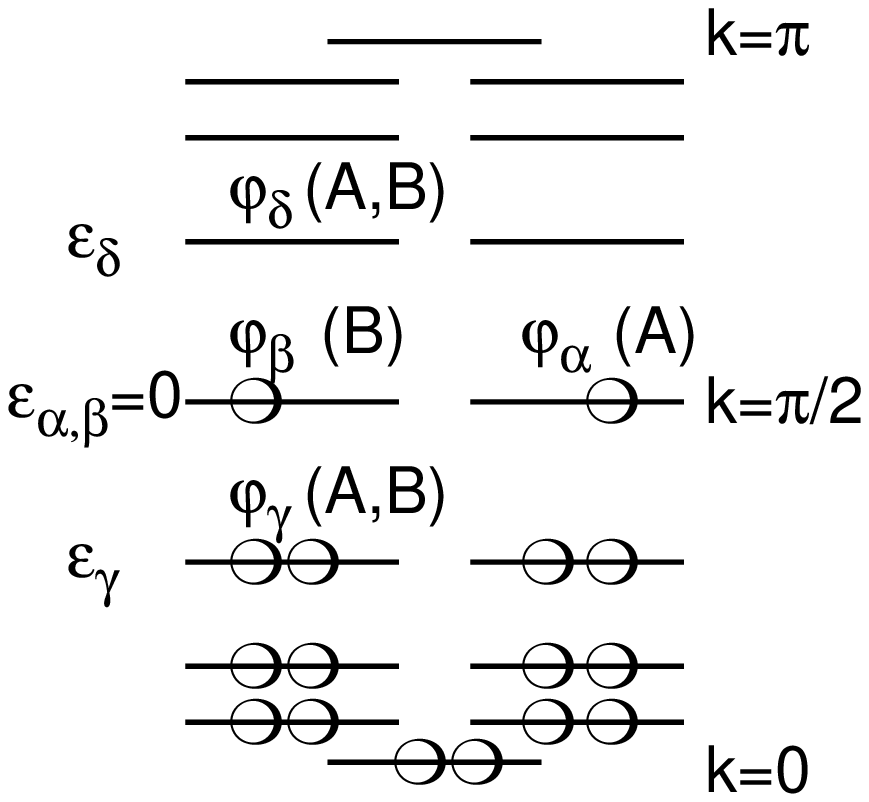}
\caption{(up) The circular molecule with $N_s=16$.
(down) The single particle eigenstates are represented by horizontal lines,
while the circles indicate occupied states at half-filling.
At mid-spectrum where $k=\pi/2$
states $\varphi_\alpha$ and $\varphi_\beta$ do not share any common site (disjoint) and are occupied by two electrons in the singlet state, thus breaking the first Hund rule.
For all other degenerate states, away from half-filling, the Hund rule is discussed in the text.}
\label{qring}
\end{figure}

The interaction potential between two electrons localized on the sites $n$ and $m$ with coordinates $r_n$, $r_m$
is considered within the extended Hubbard model to be given by
\begin{eqnarray}\label{vnm}
V_{nm}=\frac{V_L}{|r_n-r_m|}(1-\delta_{nm})+U_H\delta_{nm},
\end{eqnarray}
with $V_L$ the long range parameter and $U_H$ the Hubbard interaction term.
If{, say,}  $R_1$ is the distance between the nearest sites, $V_L/R_1$ and $U_H$
are measured in the energy unit $t$ of the hopping integral set equal to $1$.
A ring geometry with $N_s=16$ is depicted in Fig.\,\ref{qring}.

Previously, this model was used in \cite{Korkusinski2007, Bulka2011, Wrzesniewski2015} to investigate the interaction effect
in quantum dot molecules, in {core-shell} nanowire with corner localised electric charge \cite{Ozfidan2015, Sitek2017}, or in discretized quantum rings\cite{daday}.
It was also found to be a good approximation
to describe the electronic dynamics in planar models of
circular molecules as
cyclobutadiene \cite{Borden1975} or cyclooctatetraene \cite{Carl1991, Tohru2010} when the H\"ukel model is used.
The extended Hubbard model is also used in chemistry in the frame of
Pariser-Parr-Pople model Hamiltonian \cite{Sahoo2012}.

\section{Single particle states}

It is well known that in the absence of the interaction, the single-particle spectrum of circular molecules consist of a ladder of double degenerate states \cite{Salem1966, Jones1979, Mercero2015, Siga2005}. Here we briefly outline some characteristic properties, useful in the ensuing discussion.

Following the  notations in Ref.\,\cite{Siga2005} for a quantum ring one can express the
energy of the twice degenerate states to be $\epsilon_k = -2t\cos k$, where $k = 2\pi l/N_s$ $(l  = 1, 2, N_s/2-1)$
is the wavevector (with $R_1$ the distance between the sites set equal to unity).
The associated eigenstates are,
\begin{eqnarray}\label{vk1}
&&|\varphi_k^{(1)}\rangle= \sqrt \frac{2}{N_s} \sum_{n}  \sin n k |n \rangle, \\
\label{vk2}
&&|\varphi_k^{(2)}\rangle= \sqrt \frac{2}{N_s} \sum_{n}  \cos n k |n \rangle.
\end{eqnarray}
In Fig.\,~\ref{qring} this result is shown for $N_s=16$ sites.

For $k = \pi/2$ and $l = N_s/4$, the eigenstates identified above, located at energy $\epsilon(\pi/2) = 0$, satisfy
\begin{eqnarray}\label{disjoint}
\langle n|\varphi_{\pi/2}^{(1)}\rangle\langle\varphi_{\pi/2}^{(2)}|n\rangle=0,~\mbox{for all} ~n\in[1,N_s],
\end{eqnarray}
which means that they do not share any common site.
If we think of the quantum ring as a bipartite lattice with $A$ sites for $n$ odd and $B$ sites for $n$ even as sketched in Fig.\,\ref{qring},
it follows that
$|\varphi_{\pi/2}^{(1)}\rangle$ has only $A$ sites localization, and
$|\varphi_{\pi/2}^{(2)}\rangle$ has only $B$ sites localization. Such non-overlapping states are also present in the flat band of a Lieb lattice \cite{Nita1}
or as different localized edge states in 2D materials \cite{Grafena1, Grafena2, Bogdan1}. In the frame of the molecular H\"uckel model used in the field of quantum chemistry, this situation defines the disjoint non-bonding orbitals \cite{Atkins, Borden1976, Borden1994} which have relevance for spin properties at half-filling.

For wave numbers $k=0$ and $k = \pi$ ($l=0,N_s/2$) the double degeneracy of the spectrum is lifted as the single particle energies are $\epsilon_0=-2t$ and $\epsilon_\pi=+2t$, respectively.
In this case $|\varphi^{(1)}_k\rangle$ is zero, while $|\varphi^{(2)}_k\rangle$ (Eq.\,\ref{vk2}) is the only good eigenstate, which upon normalization becomes,
\begin{eqnarray}\label{vk3}
&&|\varphi_0^{(2)}\rangle=    \sqrt    \frac{1}{N_s} \sum_{n} | n \rangle, \\
\label{vk4}
&&|\varphi_{\pi}^{(2)}\rangle= \sqrt \frac{1}{N_s} \sum_{n} (-1)^n | n \rangle.
\end{eqnarray}

\section{The interacting ground state}\label{igs}

In this section, we investigate the applicability of the HFR for a pair of electrons that occupy the top states in the single particle spectrum identified above when several electronic occupancies are realized in a molecule with $N_s$ sites in the presence of the Coulomb interaction. As such, we evaluate only the differences between the
 {lowest} energies of the interacting system when a pair of electron spins form a singlet state, corresponding to total spin momentum $S = 0$ or a triplet with $S = 1$. With $E_0$ and $E_1$
 {denoting}
 the
  lowest energies in the spin sectors $S  = 0$ and $S = 1$ respectively,
we define the magnetic energy or the singlet-triplet gap as:
\begin{eqnarray}\label{deltae1}
\Delta E=E_0-E_1.
\end{eqnarray}

In a perturbative approach, $\Delta E$ is obtained, in a first order approximation, to be equal
to the exchange energy associated with the Coulomb interaction between the parallel spins in the triplet configuration
\cite{Tolea2016}. If the exchange energy is zero, a second order calculation in the Coulomb interaction is performed.

In the eigenfunction representation Eqs.\,\ref{vk1} and \ref{vk2}, the Coulomb matrix element for any four single particle quantum states, $\varphi_\alpha$, $\varphi_\beta$,
$\varphi_\gamma$ and $\varphi_\delta$, is written as $V_{\alpha \beta , \gamma \delta}$,
\begin{eqnarray}\label{vcoulombian}
V_{\alpha \beta , \gamma \delta}
=\sum _{n_1,n_2} \varphi_{\alpha}(n_1)^{\star} \varphi_{\beta}(n_2)^{\star}
                       V_{n_1 n_2} \varphi_{\gamma}(n_1)         \varphi_{\delta}(n_2),
\end{eqnarray}
with $n_1$ and $n_2$ counting the positions from $1$ to $N_s$ of the two electrons in the system and
$V_{n_1,n_2}$ from Eq.~(\ref{vnm}).

First, we consider the case of a quantum molecule with $N_e$ number of electrons that fill the energy levels with every $k \le k_0$,
with the last two electrons occupying the degenerate states $\varphi_{k_0}^{(1)}$ and $\varphi_{k_0}^{(2)}$ with a given
${k_0}\ne 0, \pi/2$ or $\pi$.
{ Actually $k_0=0$ or $\pi$ correspond to the lowest and highest non-degenerate levels,
not of interest for our discussion,
and $k_0=\pi /2$ is discussed later in this Section.}

{The singlet-triplet gap $\Delta E$ is equal to
the exchange energy} $V_{\alpha\beta,\beta\alpha}$, where $\varphi_\alpha=\varphi_{k_0}^{(1)}$ and $\varphi_\beta=\varphi_{k_0}^{(2)}$,
\cite{Tolea2016}:
\begin{eqnarray}\label{eex1}
\Delta E = 2 V_{\alpha\beta,\beta\alpha}.
\end{eqnarray}
From Eqs.\,(\ref{vk1}), (\ref{vk2}) and (\ref{vcoulombian}),
\begin{eqnarray}\label{eex2}
&&V_{\alpha\beta,\beta\alpha} = \left( \frac{2}{N_s} \right)^2 \times \nonumber\\
&&\sum_{n_1,n_2=1}^{N_s} \sin n_1k_0 \cos n_2k_0 V_{n_1n_2} \cos n_1k_0 \sin n_2k_0,~~
\end{eqnarray}
which can be written as a difference of two terms:
\begin{eqnarray}\label{eex3}
& & V_{\alpha\beta,\beta\alpha} = \frac{1}{2N_s^2}  \sum_{n_1,n_2=1}^{N_s} \cos 2(n_1-n_2)k_0 V_{n_1n_2}\nonumber\\
&- &
        \frac{1}{2N_s^2}  \sum_{n_1,n_2=1}^{N_s} \cos 2(n_1+n_2)k_0 V_{n_1n_2}.
\end{eqnarray}
The first term in Eq.\,\ref{eex3} is just the Fourier transform $\frac{1}{2}V(2k_0)$ (see also Appendix).
 The second term can be shown to {\it vanish}
for all the allowed values of the wave vector $k_0=2\pi l/N_s$, except for $k_0=0,\pi /2$ and $\pi$.
For $k_0=\pi /2$, should such a value exist in spectrum (for $4N$ molecules), the second term cancels exactly the first one and an evaluation in the second order of the Coulomb interaction is needed.
%
Leaving this single exception aside for the moment, the spin splitting energy (Eq.\,\ref{eex1}) is therefore given by the Fourier transform for the wave number $2k_0$:
\begin{eqnarray}\label{deltaei}
\Delta E=V(2k_0).
\end{eqnarray}
The above equation is one of the main formal results of our paper.
Consequently, the sign of $V(2k_0)$ determines the spin configuration in the ground state, since $\Delta E > 0$ means a triplet ground state and $\Delta E < 0$ a singlet ground state.

Since the sign of the function $V(q)$ dictates the spin of the ground state, a legitimate question is wether negative values can be obtained - i.e. singlet ground state and the anti-Hund rule, or is $V(q)$ always positive.
The minimum value of $V(q)$ can be inferred by formally considering $q$ as a {\it continuous} variable in Eq.\,\ref{vspec2} which upon differentiation generates  $q_{\min} = \pi$ \cite{comment}.

An interesting analytical result is obtained from the large $N_s$ limit. We can consider $N_s=even$, since only for even number of sites the wave vector $k_0=\pi /2$ is a physically allowed value.
One can show that, for $N_s=even$ (see Appendix):

\begin{equation}\label{limvq}
V(\pi)\rightarrow \frac 1{N_s}\big[{U_H}-2 \ln 2\frac{V_L}{\Delta}\big], ~ \text{for}~ N_s \rightarrow \infty, \Delta = 2\pi R/N_s.
\end{equation}

This implies that a given Fourier component like $V(2k_0)$ is always positive for any
$U_H> 2 \ln 2\frac{V_L}{\Delta}\simeq 1.386 \frac{V_L}{\Delta}  $. This represents, for instance, reasonable values
of the Hubbard and long range parameters ratio for an artificial quantum dot arrays { model used in} \cite{Tolea2016}.
In this case the spin energy gap from Eq.\,\ref{deltaei} is always positive ($\Delta E >0$). The examples presented in the next Section all suggest that the triplet is ground state for physically reasonable reasons. However, as Eq.~(\ref{deltaei}) indicates, mathematically situations with a preferred singlet ground state are possible. To separate the two instances, it is insightful to plot the long-range part of $V(q)$, i.e. for $U_H=0$.

\begin{figure}
\centering
\includegraphics[scale=0.7]{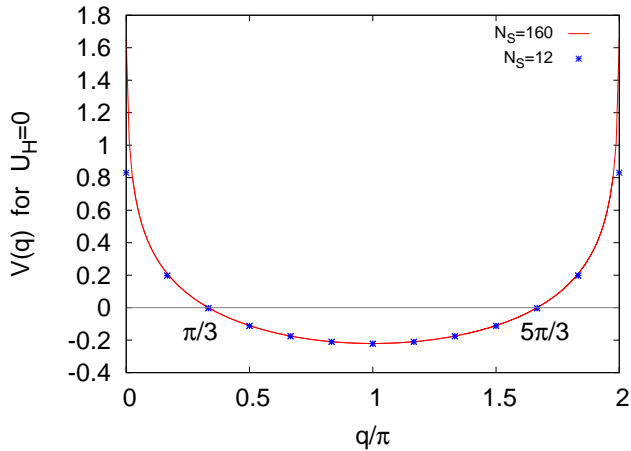}
\caption{Long-range part of V(q) (constants disregarded i.e. $R=1$ and $V_L=1$) for $N_s=12$ and $N_s=160$. The physically allowed values for the wave number are
multiples of $\frac{2\pi}{N_s}$. We notice $V(\pi) \to -\ln 2 / \pi$.}
\label{vdeq1}
\end{figure}

From Fig.~\ref{vdeq1}, one notices that for $q<\pi /3$ or $q>5\pi /3$, $V(q)>0$  while for the middle interval $q\in (\pi /3,5\pi /3)$ the long range part of $V(q)$ takes negative values, which are to be compared with the on-site Hubbard part (which is always positive) in order to decide the sign of $\Delta E$.

In conclusion when the double degenerate states with the wave number  $k\neq \pi/2$
are occupied with the last two electrons, the ground state is decided by the sign of the exchange energy in the triplet configuration.
 The triplet is always the ground state if $k\in (0,\pi/6)$ or $k\in (5\pi /6,\pi)$,
for any values of the interaction parameters $U_H$ or $V_L$.
For intermediate values $k\in [\pi/6,5\pi /6]$, the singlet can become ground state if the long range interaction exceeds
a $k$ dependent threshold value (relatively to $U_H$).

A significant exception to this rule, as mentioned previously, occurs in the case of a molecule whose number of sites $N_s$ is a multiple of four, occupied by $N_e = N_s$. In this case the two mid spectrum states with $k_0=\pi/2$ are occupied by two electrons. The exchange energy vanishes for this case and the perturbative calculation in the Coulomb potential must be carried out in the second order to determine $\Delta E$. As we show below,
the singlet state is {\it always} the interacting ground state of the system, as $\Delta E < 0$, and an anti-Hund rule situation is obtained.

For the beginning, we consider a single particle excitation process
from one state with wave number $\pi/2-q'<k_0$ to another state with wave number $\pi/2+q>k_0$.
These states need to have
the same symmetry properties to allow single particle excitation between them.


%
With the notations
$\varphi_\alpha=\varphi_{\frac{\pi}{2}}^{(1)}$, $\varphi_\beta=\varphi_{\frac{\pi}{2}}^{(2)}$,
$\varphi_\gamma=\varphi_{\frac{\pi}{2}-q'}^{(1)}$ and
$\varphi_\delta=\varphi_{\frac{\pi}{2}+q}^{(1)}$
we have the following formula for the spin splitting energy in the second order of perturbation \cite{Tolea2016}:
\begin{eqnarray}\label{deltae2}
\Delta E=2V_{\alpha\beta,\beta\alpha}+ \frac{4 V_{\delta\alpha,\alpha\gamma}
                            V_{\delta\beta,\beta\gamma}}{\Delta_{\delta,\gamma}},
\end{eqnarray}
where $\Delta_{\delta,\gamma}$ is the excitation energy $\Delta_{\delta,\gamma}=\epsilon_\delta-\epsilon_\gamma$.
A similar formula as Eq.\,\ref{deltae2} can be obtained with self consistent orbitals
from Ref.\,\cite{Kollmar1978} if one consider that {the singlet and triplet wave functions} are the same.

The Coulomb matrix elements
that enter in Eq.\,\ref{deltae2} are obtained straightforwardly:
\begin{eqnarray}
\label{v1}
&& V_{\alpha\beta,\beta\alpha}=0, \\
\label{v2}
&& V_{\delta\alpha,\alpha\gamma}=\frac{1}{2}\left[V(q)+V(\pi-q)\right]\delta_{q,q'},\\
\label{v3}
&& V_{\delta\beta,\beta\gamma}=-V_{\delta\alpha,\alpha\gamma},
\end{eqnarray}
for $q,q'\in (0, \pi/2)$.
The first order cancellation in Eq.\,\ref{v1} is readily obtained when using the disjointness relation
of the two states with wave number $k_0=\pi/2$ from Eq.\,\ref{disjoint}.
For Eqs.\,\ref{v2} and \ref{v3}, after further arrangements, we use the summation of the Fourier transformations from Eq.\,\ref{vspec3}.
The negative sign in Eq.\,\ref{v3} is the one that will lead to negative splitting energy and HFR violation.

If we consider now the single particle excitation between the cosine functions
$\varphi_\gamma=\varphi_{\frac{\pi}{2}-q'}^{(2)}$ and
$\varphi_\delta=\varphi_{\frac{\pi}{2}+q}^{(2)}$ we find out that
the Coulomb matrix elements $V_{\delta\alpha,\alpha\gamma}$ and $V_{\delta\beta,\beta\gamma}$ only change the sign compared to
those generated by the sine functions. The difference is that in this case $q,q'$ can have also the value $\pi/2$ corresponding to the
transition between the two extreme energy states from Eqs.\ref{vk3} and \ref{vk4}, {which means that $q,q'\in (0, \pi/2]$}.

We are holding now all possible single particle transition processes between the
states with wave numbers $\pi/2- q$ and $\pi/2+ q$ for any possible value of $q$.
Using above considerations
in Eq.\,\ref{deltae2} and summing  the terms for all pairs of single particle states $\varphi_\gamma$, $\varphi_\delta$
we obtain the following relation for the spin energies splitting:
\begin{eqnarray}\label{deltaeii}
\Delta E=-\frac{2(V(\pi/2))^2}{|\epsilon_{0}|}-\sum_q\frac{(V(q)+V(\pi-q))^2 }{ |\epsilon_{\frac \pi 2 - q}|}
\end{eqnarray}
with $\epsilon_{\pi/2-q}$ and $\epsilon_{0}$ the single particle energies.

The first term of the above equation accounts for excitations from the lowest non-degenerate state ($k_0=0$)
to the highest one ($k_0=\pi$), which is also non-degenerate. In the case of the {\it four} atom molecule, it is the only term existent.
For all the other $4N$ molecules with $N\ge 2$, the second term must be considered as well,
taking account for the allowed excitations between double degenerate states symmetrically placed below and above the mid-spectrum.
The summation is over the values $q= {\frac{2\pi}{N_s}, \frac{4\pi}{N_s},\cdots,  \frac {(N_s-4)\pi}{2N_s}}$.
Eq.\,\ref{deltaeii} therefore shows a negative sign of the spin splitting energy (i.e. $\Delta E < 0$) and therefore a singlet ground state and anti-Hund situation for the half-filled $4N$ molecule.

\section{Examples}

In this Section, we show calculations of the singlet-triplet level spacing for some simple molecules (either made of atoms or of quantum dots), using for the Hubbard or long-range interactions values or formulas proposed in literature.

Two situations when the Hund or anti-Hund situations are decided by the ratio between Hubbard  and the long-range interactions
are shown in Table\,{I}
for a triangle molecule and for an octagon molecule at various filling factors except for the half-filling.

\begin{table}[h!]
\label{table2}
\begin{center}
\begin{tabular}{|l|l|l|}
\hline
~System &  Electron Configuration & $\Delta E =E_S-E_T$ \\
\hline
& & \\
~\includegraphics[width=0.06\textwidth, height=10mm]{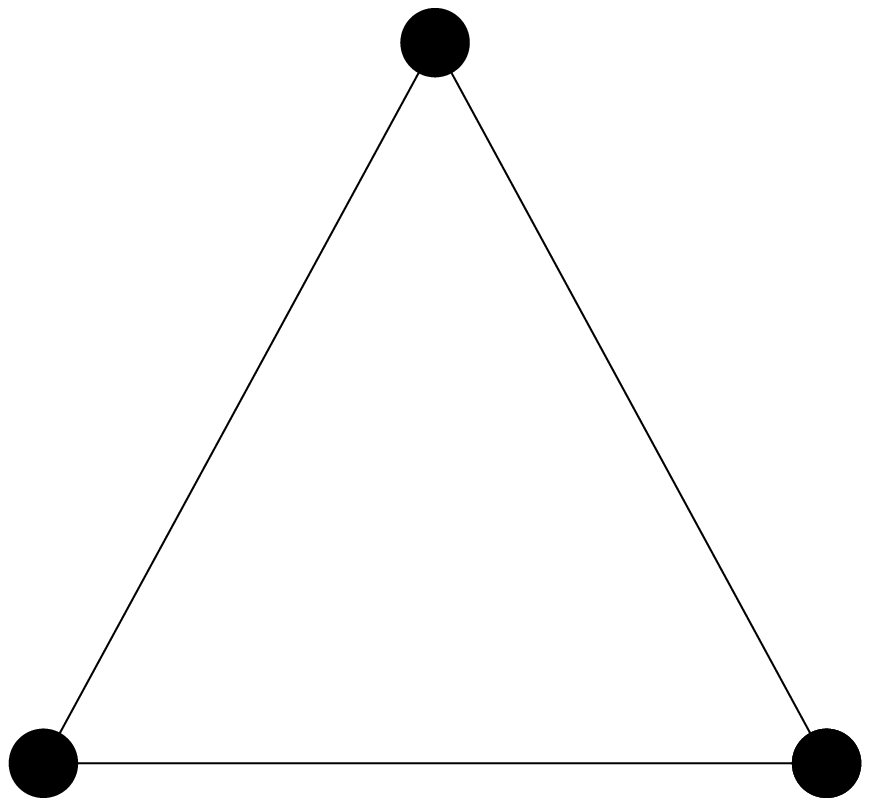} ~&
~
\includegraphics[width=0.1\textwidth, height=14mm]{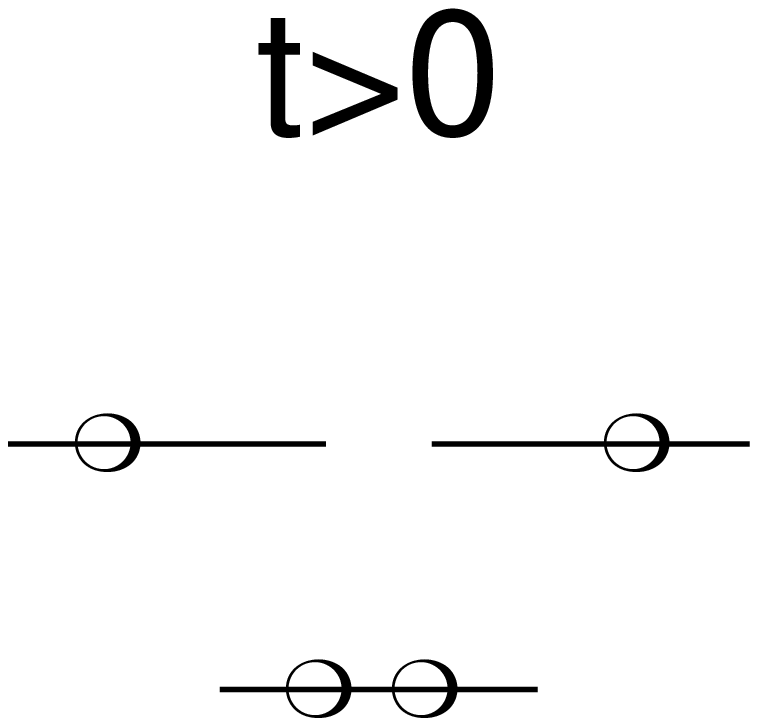} ~~
\includegraphics[width=0.1\textwidth, height=14mm]{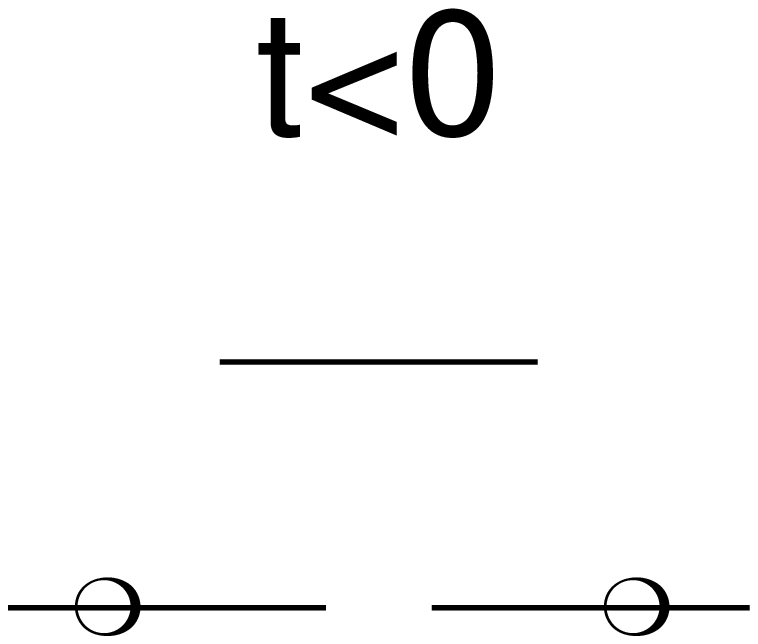} &

$ \begin{aligned}  \Delta E=~~~~~~~~~~&&\\  \frac{1}{3} \left(U_H-V(R_1)   \right)&& \\ &&\end{aligned}$ \\
\hline
  &  &  \\
~\includegraphics[width=0.06\textwidth, height=10mm]{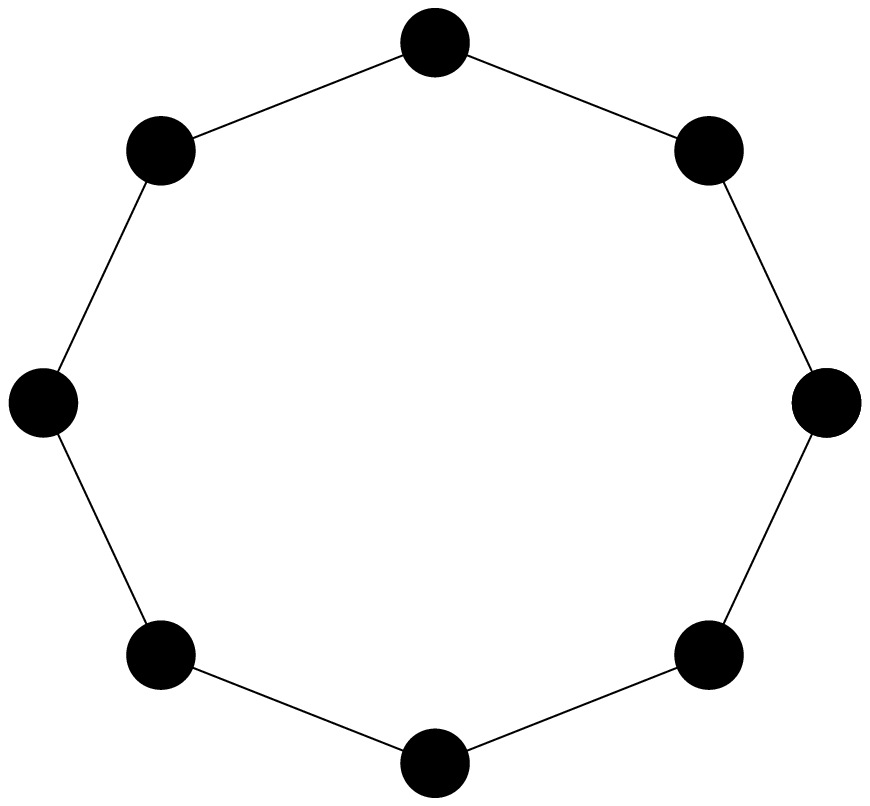} &
~
\includegraphics[width=0.1\textwidth, height=14mm]{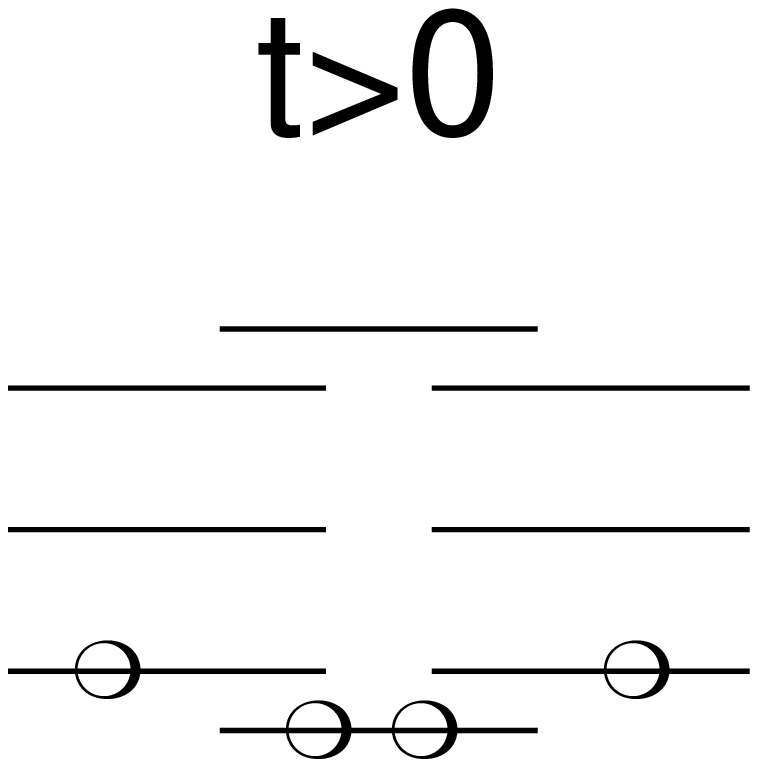}  ~~
\includegraphics[width=0.1\textwidth, height=14mm]{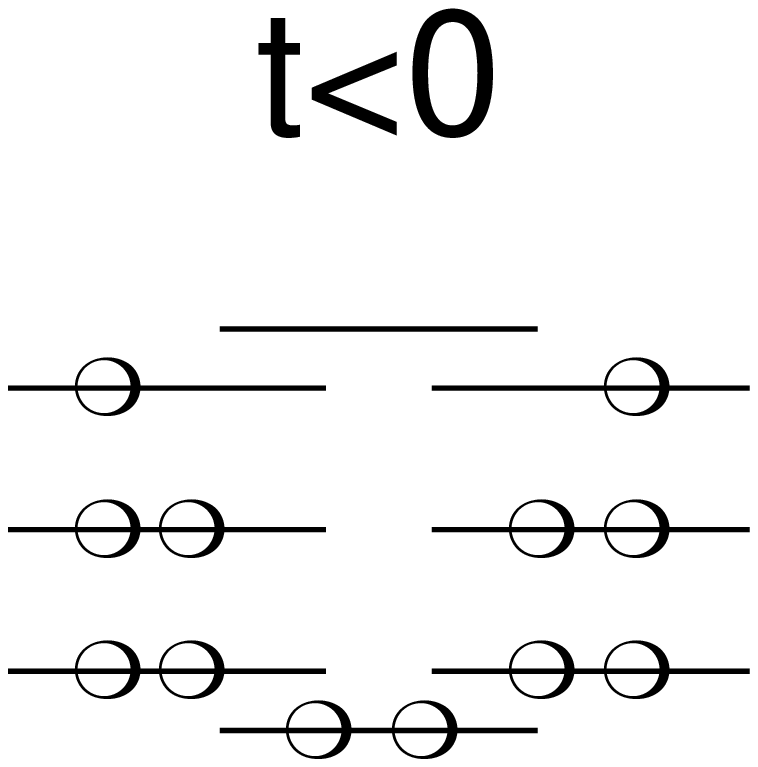}  &
$\begin{aligned}  \Delta E=~~~~~~~~~~ &&\\  \frac{1}{8} \left(U_H + V(R_4)\right) &&\\  -\frac{1}{4} V(R_2) ~~~~~~~~   && \end{aligned}$
\\
\hline
\end{tabular}
\end{center}
\caption{Singlet-triplet splitting when the last two degenerate occupied orbitals are {away of half-filling}.
The calculations are done in the first order approximation.}

\end{table}

For an artificial molecule constructed with quantum dots, one may use for instance the dot confinement model described
in \cite{Tamura2002} where the authors calculate the interaction parameters as:
\begin{eqnarray}\label{uh1}
&& U_H=\frac{e^2}{2\sqrt {2\pi}\epsilon d}, \\
\label{vl1}
&& V(R_n)=\frac{e^2}{4 \pi \epsilon R_n},
\end{eqnarray}
with { $d$ the dot diameter and $R_n$ the inter-dot distance of order $n$.
One can modify
the $V(R_n)$ and $U_H$ by varying either the dots diameter or inter-dot distances.}
Using Eqs.\,\ref{uh1} and \ref{vl1} the energy splitting for triangle molecule in Table\,I is always positive
for the dot confinement $d < R_1\sqrt {2\pi}\simeq 2.5 R_1$ {which is true in this case}.
We mention also that in \cite{Grifoni2016} the exact results for a triangle are calculated and our results
are recovered in the limit $U_H-V(R_1)\ll |t|$.

{As a matter of fact, using the condition from Eq.\,\ref{limvq}, it is easy to show that
the triplet state is always the ground state away from half-filling (as situations in Table\,I)
for the dot diameter $d<R_1$ when considerring the model in Eqs.\,\ref{uh1} and \ref{vl1}.}

The above values for $U_H$ and $V(R_n)$ may also be used to compute,
for instance, the singlet-triplet splitting in half-filled $4N$ molecules,
when one has always an anti-Hund rule. For the square and the octagon,
the results are given in Table\,II.

\begin{table}[h!]
\label{table1}
\begin{center}
\begin{tabular}{|l|l|l|}
\hline
~System  &   $\begin{aligned} \text{Electron}\\ \text{configuration} \end{aligned}$
 & $\begin{aligned} &&\text{Singlet-Triplet Splitting} \\ &&\Delta E =E_S-E_T ~~~~~ \end{aligned} $ \\
\hline

~ \includegraphics[width=0.06\textwidth, height=10mm]{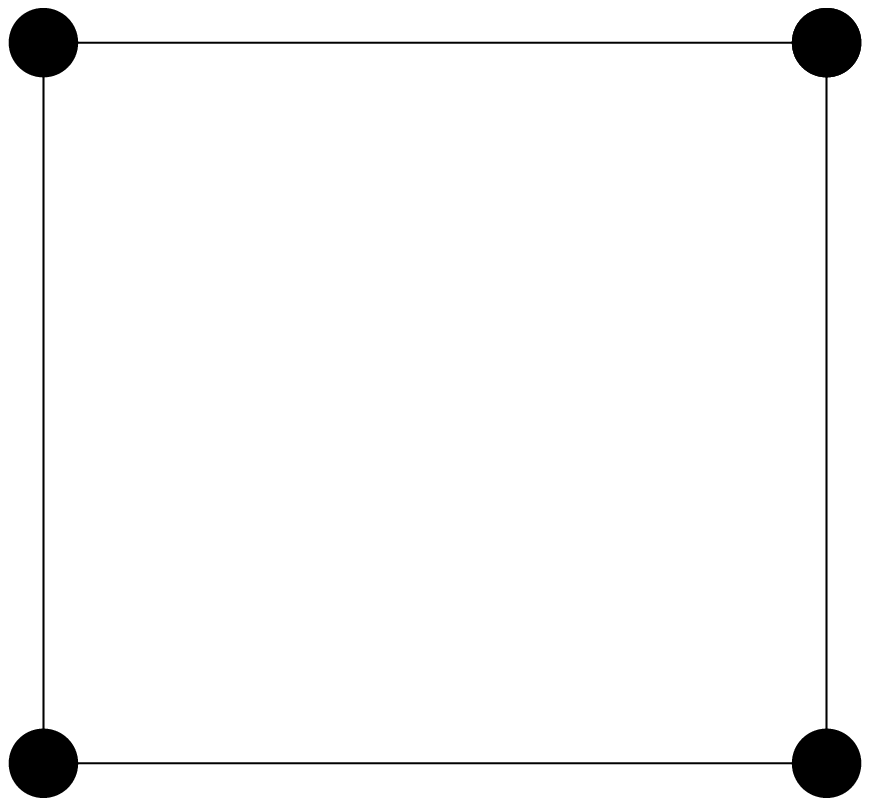} ~ &
~ \includegraphics[width=0.1\textwidth, height=14mm]{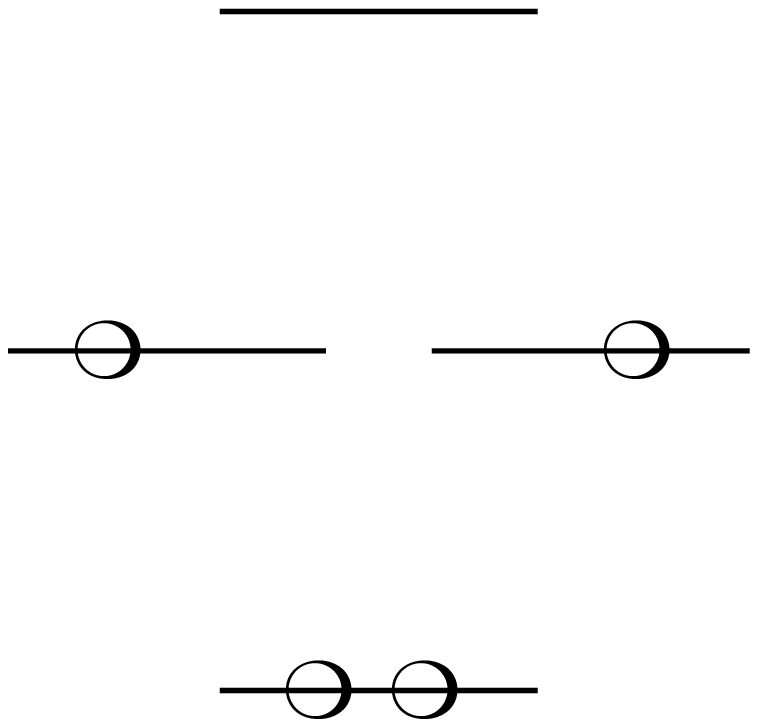} &
$\Delta E=-\frac{1}{16 |t|}  \left(U_H-V(R_2)\right)^2 $ \\
\hline

~ \includegraphics[width=0.06\textwidth, height=10mm]{octogonsilctn.eps} &
~ \includegraphics[width=0.1\textwidth, height=14mm]{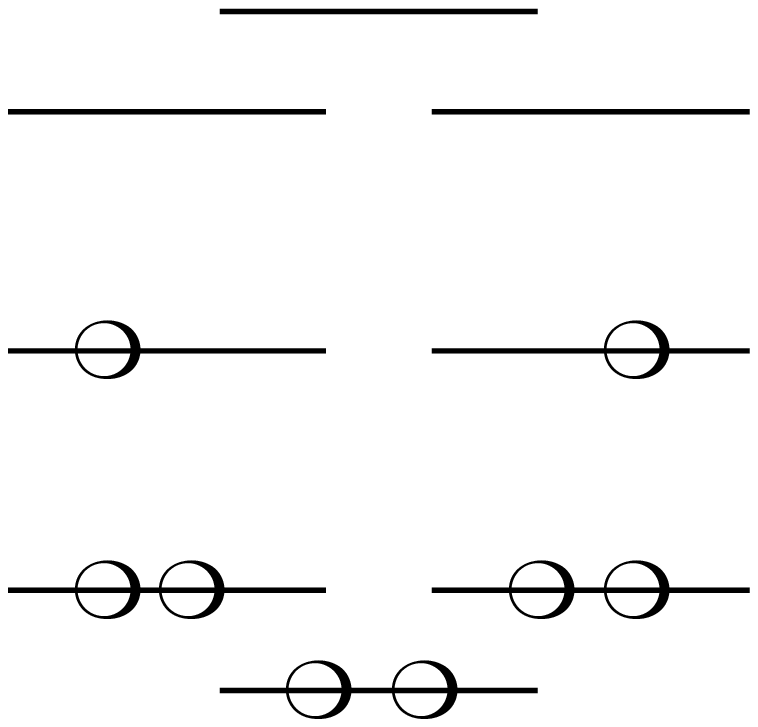} &
$\begin {aligned} \Delta E=-\frac{1}{16\sqrt 2 |t|}  \left(U_H-V(R_4)\right)^2 \\
         -\frac{1}{64 |t|}  \left(U_H+V(R_4)-2V(R_2)\right)^2    \end{aligned} $
\\
\hline
\end{tabular}
\end{center}
\caption{Negative singlet-triplet splitting for the half-filled square and octagon molecules.}
\end{table}

We now present results obtained by the formula of $\Delta E$ from Table\,II to approximate the
singlet-triplet energy splitting in the case of chemical molecules.
As an example we consider
the square model of cyclobutadiene molecule which has $N_s=4$ Carbon atoms.
We use standard parameters of hydrocarbures from \cite{Anusooya1998, Sahoo2012}, $U_H=11.26$eV, hopping energy $t=2.4eV$,
and distance between atoms $R_1=1.44\AA$.
The long range interaction is calculated now with Pariser-Parr-Pople model Hamiltonian \cite{PPP1953}
and Ohno formula \cite{Sahoo2012}
\begin{eqnarray}\label{vl2}
V(R_n)=\frac{14.397}{\sqrt{\left( \frac{14.397}{U_H} \right)^2 + R_n^2}},
\end{eqnarray}
with $V(R_n)$ and $U_H$ in eV and $R_n$ in $\AA$.
From the formula of Table\,I we obtain the value $\Delta E=-0.72$eV for a square model of cyclobutadiene,
close to numerical values $\Delta E_{SP}=-0.71$eV obtained in \cite{Kollmar1978} considerring the spin polarization effect
and minimal base for molecular calculation.

For the planar model of cyclooctatetraene molecule with $N_s=8$ carbon atoms, if we keep the same parameters as above,
except $R_1=1.40\AA$, we obtain $\Delta E=-0.88$eV
that is lower than other values reported in literature, but comparable as $\Delta E_{ST}=-0.68$eV or $-0.34$eV in \cite{Hrovat1992}.

 The perturbative approach described in this paper, although valid only for {\it small} values of the interacting potential, offers a good qualitative description of the spin configuration of the ground state in the case of circular molecules with a large number of atoms. Full analytical results in the presence of interaction are only available for the smallest $N=3$ molecule \cite{Grifoni2016}, while exact diagonalization numerical results are already somewhat computationally demanding even for the $N=8$ (see \cite{Tolea2016}). Such methods are not at all feasible for large $N$.
In a direct comparison, for the smaller circular molecule with $N=3$, our perturbative result from Table 1 differs from the exact result \cite{Grifoni2016} by less than 2 \% for $U_H-V(R_1)<0.1t$ and less than 6\% for $U_H-V(R_1)<0.3t$. In \cite{Tolea2016}, for $N=8$, if only the on-site Hubbard interaction is considered, we find good correspondence between the perturbative results and the exact diagonalization ones for $U_H<t$.

\section{Conclusion}


In this paper we have studied the first Hund rule
in circular molecules, for cases when the two most energetic electrons occupy a pair of degenerate levels.
The quantity of interest is the singlet-triplet energy gap $\Delta E$, which was expressed
in terms of the Fourier transform of
the interacting potential. Both on-site ($U_H$) and long range ($V_L$) interactions have been considered within an extended Hubbard model.

A special case is found for the $4N$ molecule at half filling, for which the first order energy correction (i.e. the exchange energy) vanishes and the second order gives {\it always} the singlet as ground state, and thus an anti-Hund situation. Since the $4N$ molecule is a bipartite lattice, we find ourselves in the frame of the Lieb theorem \cite{Lieb}, but with a more complex potential including arbitrary long range interaction.

For all the other cases, the exchange energy does not vanish and its sign decides the ground state. Our results show that,
 depending on the total number of electrons in the system (i.e. the wave number $k_0$ of the highest occupied levels)
we meet the two distinct situations. A triplet ground state is realized for {\it any} values of interaction parameters
 if $k_0\in (0,\pi/6)$ or $k_0\in (5\pi/6,\pi)$. On the other hand, for $k_0\in [\pi/6,\pi/2) \cup (\pi/2,5\pi/6]$  the singlet ground state is mathematically possible, with the highest probability for
$k_0$ close to $\pi/2$.
A necessary condition for singlet ground state around $k_0=\pi/2$ is
$V_L/\Delta>U_H/(2 \ln 2)$ ($\Delta$ is the nearest neighbors distance measured on the circle, i.e. $\Delta=2\pi R/N_s$).

The described formalism is applied for some few-atoms circular molecules, either real or artificial, in Section V.

The results hold for arbitrary Hubbard or long range interactions, as well as for any number of atoms in the circular molecule. Such generality is owed to the fact that the singlet-triplet level spacing was analytically expressed in terms of the Fourier transform of the interaction potential.

Apart from providing  detailed spectral calculations for molecules of potential interest,
our studies may be also relevant for understanding various origins of non-trivial spin alignment.

{\bf{Acknowledgements.}}
Thw work is supported by the Core Program grant No. PN16-480101 and National Research Program PN-III-P4-IDPCE-2016-0221.

\appendix

\section{The Fourier transform of the interaction potential [the V(k) function]}\label{fn}

The long range part of the potential $V_{nm}$ in Eq.\,\ref{vnm} depends only on the distance $R_{nm}$ between the points $n$, $m$
that, for the ring geometry, counts only the minimum number of sites from $n$ to $m$.
Then we can define the potential $V(R_n)$ for any integer $n$:
\begin{eqnarray}\label{vrn}
&& V(R_n)=\frac{V_L}{R_n}(1-\delta_{R_n,0})+U_H\delta_{R_n,0} \nonumber \\
&& ~~\text{with} ~~ R_n=2R|\sin  {\pi n } / {N_s}|,
\end{eqnarray}
with the length $R_n$ measuring the distance between two points separated by $n$ succesive sites on a circle of radius R.

The potential $V(R_n)$ from Eq.\,\ref{vrn} has the periodicity $V(R_n)=V(R_{n+N_s})$
and we define the Fourier transformation
\begin{eqnarray}\label{vdek}
V(k)=\frac{1}{N_s}\sum_{n=1}^{N_s} e^{ikn} V(R_n),
\end{eqnarray}
with the wave number  $k=\frac{2\pi}{N_s}l$ with $l$ integer.

In the calculation of ground state properties from Section\,\ref{igs} we use the following properties of $V(k)$:
\begin{eqnarray}\label{vspec1}
 V(k)&&=V(k)^\star,\\
\label{vspec2}
 V(k)&&=\frac{1}{N_s}\sum_{n=1}^{N_s} \cos {kn} V(R_n) \nonumber \\
&& =\frac{U_H}{N_s}+
\frac{V_L}{N_s}\sum_{n=1}^{N_s-1} \frac{\cos kn }{R_n}.
\end{eqnarray}
The Eq.\,\ref{vspec1} is immediate using $V(R_n)=V(R_{N_s-n})$ in Eq.\,\ref{vdek} and Eq.\,\ref{vspec2} follows from
Eq.\,\ref{vspec1} using also the explicit form of the potential from Eq.\,\ref{vrn}.

Using the definition of the Fourier transform from Eq.\,\ref{vdek} we obtain the relation
\begin{eqnarray}\label{vspec3}
V(q)+V(\pi-q)=\frac{2}{N_s}\sum_{n=2(\text{even})}^{N_s} \cos qn V(R_n)
\end{eqnarray}
that is used to obtain Eqs.\,\ref{v2} and \ref{v3}.

As mentioned also in the main text, if we treat $k$ as a continuous variable, then the derivative of Eq.\ref{vspec2}
cancels for $k=\pi$, where the function has a minimum (as seen in Fig.2). Whether this minimum is negative or remains positive, depends on the ratio $V_L/U_H$. In order to calculate $V(\pi)$ in the limit of large number of sites ($N_s\rightarrow \infty$) one evaluates, up to a constant:
\begin{equation}
\lim_{N\rightarrow \infty}  \Big[ \frac\pi N \sum_{n=1}^{N-1} \frac{\cos~ \pi n }{sin~ \pi n/N}\Big]. \label{eq:sum}
\end{equation}
This is done by taking into account that,
\begin{equation}
\lim_{N\rightarrow \infty} \sum_{n=1}^{N}\frac{(-1)^n}{n}=-\ln 2\;, \label{eq:sum-exact}
\end{equation}
an equality that reproduces the first terms in Eq.\,\ref{eq:sum}, since
for small $n$, $cos~ n\pi = (-1)^n$ while $sin~ \pi n/N \simeq \pi n/N$
in the limit $N\rightarrow \infty$.
Terms calculated for intermediate values of $n$ generate vanishing contributions.
The last term in Eq.\,\ref{eq:sum} ($n\rightarrow N$) reproduces
in magnitude terms present in Eq.\,\ref{eq:sum-exact}, as $sin~ (\pi - a)=sin~a$.
Whether the terms are reproduced with the same sign or opposite one is decided by the parity of $N$.

As a result, the limit of Eq.\,\ref{eq:sum} is $-2 \ln 2$ for $N=even$ and $0$ for $N=odd$. This proves Eq.\,\ref{limvq}.



\begin{thebibliography}{100}
\bibitem{comment}{For the present discussion, which temporarily regards $q$ as a continuous variable, it is irrelevant that the value $q=\pi$ is precisely the exception discussed in text for which Eq.\,13 does not give the energy splitting.}
\end{thebibliography}
\end{document}